\documentclass{aa}  
\usepackage{url}
\usepackage{graphicx}
\usepackage{natbib}
\usepackage{hyperref}
\usepackage{breakurl}
\usepackage[font=small,labelfont=bf,labelsep=period,skip=2.5mm]{caption}
\usepackage{txfonts}
\begin{document}
  \title{XMM-Newton observations of the X-ray soft polar QS Telescopii
     \thanks{Based on observations obtained with XMM-Newton, an ESA science
     mission with instruments and contributions directly funded by ESA Member
     States and NASA.}
   }



   \author{I.~Traulsen  \inst{\ref{iag},\ref{aip}}
     \and  K.~Reinsch   \inst{\ref{iag}}
     \and  A.~D.~Schwope\inst{\ref{aip}}
     \and  V.~Burwitz   \inst{\ref{mpe}}
     \and  S.~Dreizler  \inst{\ref{iag}}
     \and  R.~Schwarz   \inst{\ref{aip}}
     \and  F.~M.~Walter \inst{\ref{suny}}
          }

   \offprints{I.~Traulsen}

   \institute{Institut f\"ur Astrophysik, Georg-August-Universit\"at
              G\"ottingen, Friedrich-Hund-Platz 1, 37077 G\"ottingen,
              Germany\\
              \email{traulsen@astro.physik.uni-goettingen.de}\label{iag}
         \and
              Astrophysikalisches Institut Potsdam, An der Sternwarte 16, 14482
              Potsdam, Germany\label{aip}
         \and
	      Max-Planck-Institut f\"ur Extraterrestrische Physik, P.O.~Box
              1312, 85741 Garching, Germany\label{mpe}
         \and
              Department of Physics and Astronomy, Stony Brook University,
              Stony Brook, NY 11794-3800, USA\label{suny}
             }

   \date{Received 17 December 2010; accepted 15 March 2011}

%
%

  \abstract
   {On the basis of XMM-Newton observations, we investigate the energy balance
     of selected magnetic cataclysmic variables, which have shown an extreme
     soft-to-hard X-ray flux ratio in the ROSAT All-Sky Survey.}
   {We intend to establish the X-ray properties of the system components,
     their flux contributions, and the accretion geometry of the X-ray soft
     polar QS~Tel. In the context of high-resolution X-ray analyses of
     magnetic cataclysmic variables, this study will contribute to better
     understanding the accretion processes on magnetic white dwarfs.}
   {During an intermediate high state of accretion of QS Tel, we have obtained
     20\,ks of XMM-Newton data, corresponding to more than two orbital periods,
     accompanied by simultaneous optical photometry and phase-resolved
     spectroscopy. We analyze the multi-wavelength spectra and light curves
     and compare them to former high- and low-state observations.}
   {Soft emission at energies below 2\,keV dominates the X-ray light
     curves. The complex double-peaked maxima are disrupted by a sharp dip in
     the very soft energy range ($0.1-0.5\,\mathrm{keV}$), where the count
     rate abruptly drops to zero. The EPIC spectra are described by a
     minimally absorbed black body at 20\,eV and two partially absorbed
     \textsc{mekal} plasma models with temperatures around 0.2 and 3\,keV. The
     black-body-like component arises from one mainly active, soft X-ray
     bright accretion region nearly facing the mass donor. Parts of the plasma
     emission might be attributed to the second, virtually inactive pole. High
     soft-to-hard X-ray flux ratios and hardness ratios demonstrate that the
     high-energy emission of QS~Tel is substantially dominated by its X-ray
     soft component.}
   {}

%
%

   \maketitle

%
%
%

%
%
%
\section{Introduction}

  \object{QS~Tel} belongs to the group of \object{AM~Her}-type cataclysmic
  variables, in which the field strength of the white-dwarf primary is high
  enough to prevent the formation of an accretion disk. It has been discovered
  in the ROSAT All-Sky Survey, independently in the soft X-ray regime by
  \citet{beuermann:93} and in the EUV by \citet{buckley:93}. Survey and
  pointed ROSAT observations revealed a distinct soft X-ray flux
  component. With an orbital period of 2.33\,hrs, it is one of the few systems
  that settle in the period gap of cataclysmic variables. The magnetic field
  strength $M=50-80\,\mathrm{MG}$ of the white dwarf in QS~Tel ranges among
  the highest values measured in AM~Her-type systems. Undergoing frequent
  changes between high and low states and occasionally switching between
  one-pole accretion \citep[e.\,g.][]{buckley:93} and two-pole accretion
  \citep[e.\,g.][]{rosen:96}, QS~Tel exhibits a variety of accretion behaviors
  and has been subject to various multi-wavelength studies.

  We are obtaining XMM-Newton data of soft X-ray selected polars which have
  not been subject to high-resolution X-ray observations before, and
  investigate the system properties and a potential dominance of the soft over
  the hard X-ray flux ('soft X-ray excess'). This paper on the soft polar
  QS~Tel is the second in a series dealing with the results of our
  campaign. In Sect.~\ref{sec:data}, we present our observational X-ray and
  optical data. Sect.~\ref{sec:photo} summarizes the results of the
  light-curve and Sect.~\ref{sec:spectra} of the spectral analyses,
  respectively. Finally, we discuss the implications on accretion state and
  accretion geometry in Sect.~\ref{sec:discuss}.

%
%
\section{Observations and data reduction}
\label{sec:data}

  QS~Tel was observed with XMM-Newton on September 30, 2006 for 20\,ks
  (archived under observation ID 0404710401). The EPIC/pn detector was
  operated in large window mode with the medium filter, EPIC/MOS2 in small
  window mode with the thin filter. EPIC/MOS1 suffered from a full scientific
  buffer in timing mode and collected too little signal. With the optical
  monitor, we performed ultraviolet fast mode photometry in the
  2050$-$2450\,{\AA} band using the UVM2 filter. We employ \textsc{sas}\,v8.0
  standard tasks for the data reduction and apply an effective area correction
  as described by \citet{traulsen:10} for the EPIC spectra of
  \object{AI~Tri}. For the EPIC instruments, we choose a circular source
  region with a radius of 25\,arcsec. The EPIC/pn background can be taken from
  the same chip as the source.  Due to gaps in the detector plane close to the
  source, we determine the EPIC/MOS2 background from a source-free region on
  an outer CCD. The net peak count rates of $2.2\pm 0.6\,\mathrm{cts\,s}^{-1}$
  for EPIC/pn and $0.70\pm 0.26\,\mathrm{cts\,s}^{-1}$ for EPIC/MOS2,
  respectively, are about a factor of eight lower than expected during high
  states from the ROSAT All-Sky Survey \citep{voges:99,schwope:95}. This means
  that QS~Tel was in an intermediate high state of accretion at the epoch of
  our observations.

  Contemporary optical observations have been performed at the CTIO
  observatory of the SMARTS consortium on October 1, 2006. $B$-band photometry
  with the ANDICAM at the 1.3\,m telescope covers 85\,\% of an orbital period
  with a cycle time of 100\,s (73 data points). Taken under poor weather
  conditions, the data are non-photometric with a low signal-to-noise. Light
  curves are derived by differential photometry and calibrated as described by
  \citet{gerke:06}. 33 phase-resolved optical spectra in the
  3500$-$5300\,{\AA} range were obtained with the R-C spectrograph at the
  1.5\,m telescope during about 80\,\% of an orbital cycle. They show a
  spectral resolution of 4.5\,{\AA} in FWHM and a time resolution of
  214\,s. About half of the spectra are affected by clouds and cannot be used
  for a quantitative spectral analysis.

  All times have been corrected to the barycenter of the solar system using
  the JPL ephemeris \citep{JPL} and converted to terrestrial time.

  \begin{figure}
    \centering
    \includegraphics[width=8.8cm]{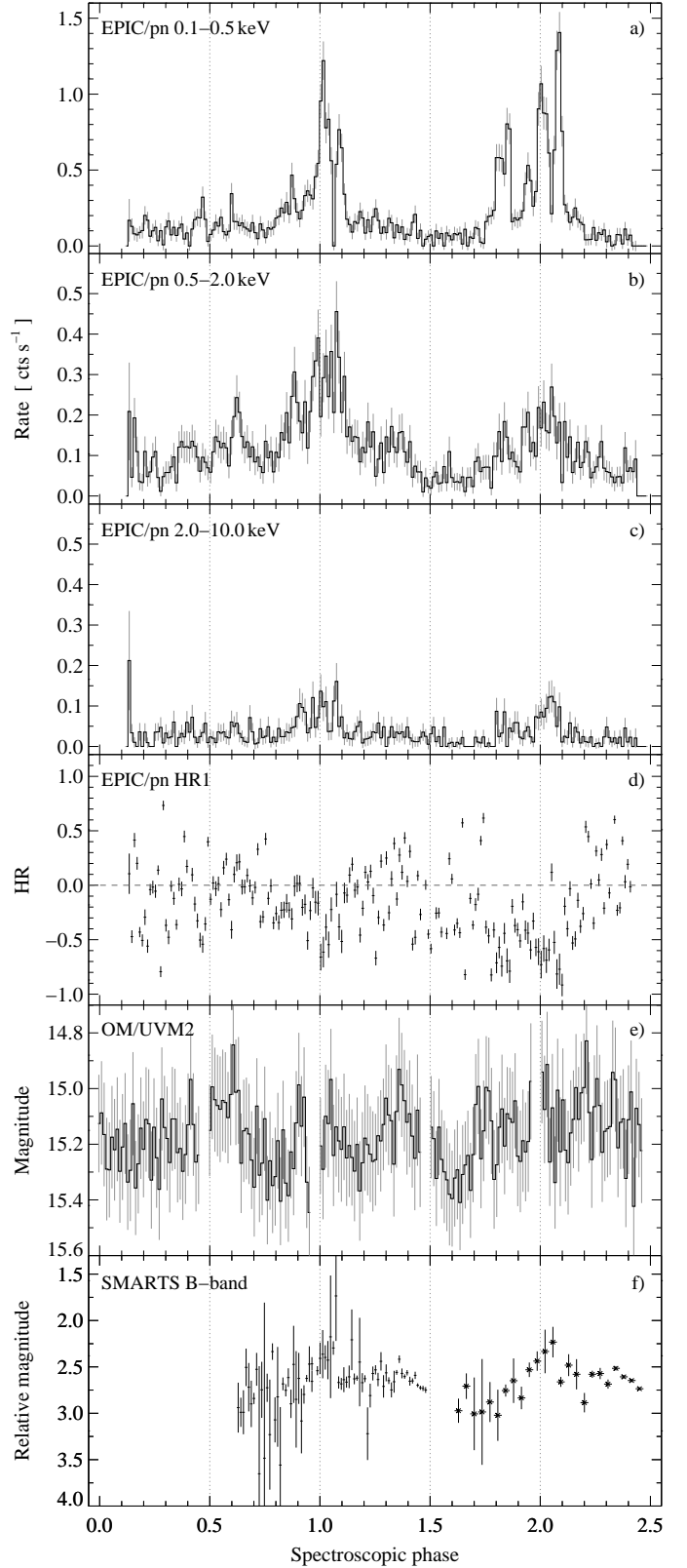}
    \caption{Multi-wavelength light curves of QS~Tel, folded on the orbital
      period. Phase zero corresponds to the inferior conjunction of the
      secondary star. \textbf{a)$-$c)} EPIC/pn light curves. \textbf{d)}
      Hardness ratio HR1$=$(H$-$S)/(H$+$S) between the counts in the
      0.1$-$0.5\,keV and in the 0.5$-$2.0\,keV ranges.  \textbf{e)}
      Ultraviolet OM light curve in the 2050$-$2450\,{\AA} band. All
      XMM-Newton data are shown in time bins of 100\,s. \textbf{f)} Optical
      SMARTS light curve, obtained simultaneously to the XMM-Newton
      observation. The data have been plotted twice, to the left with the
      original cycle time of 100\,s, to the right rebinned into intervals of
      300\,s.}
    \label{fig:multilc}%
  \end{figure}

%
%
\section{Optical data and multi-wavelength light curves}
\label{sec:photo}

\subsection{Constraining the binary ephemeris}

  The optical light curve profiles of QS~Tel exhibit a changing morphology
  \citep[see e.\,g.][]{rosen:01} and lack a definite recurring feature that
  could be used to establish photometric ephemeris. \citet{schwope:95} derive
  a linear spectroscopic ephemeris for the broad and for the narrow emission
  line components, defining $\varphi_\mathrm{NEL} = 0.0$ at the blue-to-red
  zero crossing (inferior conjunction of the secondary). In order to verify
  the accuracy of their ephemeris, we investigate new optical phase-resolved
  spectra of QS~Tel ($\Delta\varphi = 0.025$).  The radial velocity curves
  show a large velocity amplitude $K > 300\,\mathrm{km\,s}^{-1}$ and an
  asymmetric, non-sinusoidal shape, as described by \citet{buckley:93},
  \citet{schwope:95}, and \citet{rosen:96}. Since our radial-velocity
  measurements are consistent with their data to better than $\Delta\varphi
  \sim 0.06$, we consider the ephemeris given by \citet{schwope:95} to be
  still valid for our 2006 September 30 observations and refer to the phasing
  of the narrow emission line component (their Eq.~2) throughout the analysis.
  We find no clear evidence of an orbital period change.

\subsection{X-ray light curves}
\label{sec:xray}

  Figure~\ref{fig:multilc}a$-$d show the EPIC/pn photometric data folded on
  the orbital period. The soft X-ray emission at energies between 0.1 and
  0.5\,keV is characterized by high flaring variability and a distinct
  bright/faint behavior. At energies between 0.5 and 2.0\,keV, the count rate
  is lower, and the bright phases are less pronounced than in the softest
  energy range. The flux at energies above 2\,keV is very low, but non-zero
  during almost the complete orbital cycle. In all energy bands, the bright
  phases last about 40\,\% of the binary revolution. The maxima are two-fold
  disrupted by dips in the flux: A broader, shallow dip lasting $\Delta\varphi
  \sim 0.08$ occurs around orbital phase 0.9. A sharp dip at $\varphi \sim
  0.07$ ($\Delta\varphi < 0.02$) appears to be restricted to the very soft
  X-ray component at energies between 0.1 and 0.5\,keV. During this short dip
  phase, the count rate nearly drops to zero (cf.\ Fig.~\ref{fig:brightlc}),
  and the hardness ratio increases. We discuss the possible origin of the dips
  in Sect.~\ref{sec:dips}.

\subsection[B-band and ultraviolet light curves]{$B$-band and ultraviolet light curves}
\label{sec:opt}

  The ultraviolet light curve at an effective wavelength of 2310\,{\AA}
  (Fig.~\ref{fig:multilc}e) has a small amplitude of about $0.2\,\mathrm{mag}$
  and does not repeat itself during the two orbital cycles covered. The
  optical $B$-band light curve also shows little modulation with the orbital
  phase and resembles the $V\!$-band light curves during a low state of
  accretion obtained by \citet{ferrario:94}.
  Higher variability may be revealed after rebinning the optical data into
  intervals of 300\,s (Fig.~\ref{fig:multilc}f). Former optical light curves
  are characterized by significantly stronger orbital modulations than our
  data, show various different light-curve shapes and a striking short- and
  long-term variability. We find no obvious analog to our $B$-band and UV
  light curves in the numerous CTIO and SAAO observations presented by
  \citet{rosen:96,rosen:01}, or in the ultraviolet HST/FOS continuum light
  curves of \citet{demartino:98} and \citet{rosen:01}. Detailed descriptions
  of the different light curve profiles of QS~Tel are given by
  \citet{rosen:96,rosen:01}, and an overview of its $B$-band magnitude
  long-term evolution since 2003 by \citet{gerke:06}.

\begin{table*}
    \caption{Best-fit parameters of the models with one and two \textsc{mekal}
      plasma components, respectively, for the EPIC spectra of QS\,Tel
      (orbital mean).}
    \label{tab:xspecfits} 
    \centering
    \begin{tabular}{*{11}{c}}
      \hline\hline\\[-2.2ex]

      & $\chi^2_\mathrm{red}$ & $N_\mathrm{H, ISM}$ & $kT_\mathrm{bb}$ &
      $N_\mathrm{H, intr}$ & cover. & $kT_{\textsc{mekal}}$ &
      $kT_{\textsc{mekal}}$ & abund. & $F_\mathrm{bol}(\textsc{bbody})$ &
      $F_\mathrm{bol}(\textsc{mekal})$ \\

      & & [$10^{19}\,\mathrm{cm}^{-2}$] & [$\mathrm{eV}$] & [$10^{22}\,{\rm
          cm}^{-2}$] & & [$\mathrm{keV}$] & [$\mathrm{keV}$] & &
      [erg\,cm$^{-2}$\,s$^{-1}$] & [erg\,cm$^{-2}$\,s$^{-1}$]
      \\[0.3ex]

      \hline\\[-1.5ex]

      \multicolumn{11}{l}{\textsc{tbnew(bbody+pcfabs(mekal))}}
      \\[.4ex]
      & 1.22 & $3.3^{+5.9~}_{-2.9~}$ & $21.1^{+3.5}_{-3.0}$ &
      $16.1^{+7.7}_{-4.5}$ & $0.74^{+0.07}_{-0.11}$ & & $2.2^{+0.4}_{-0.3}$ &
      $0.4^{+0.2}_{-0.1}$ & $1.2^{+7.3}_{-0.8}\times10^{-11}$ &
      $1.7^{+0.5}_{-0.5}\times10^{-12}$
      \\[.8ex]

      \multicolumn{11}{l}{\textsc{tbnew(bbody+pcfabs(mekal+mekal))}}
      \\[0.4ex]
      & 1.06 & $9.3^{+14.7}_{-7.5}$ & $19.2^{+5.4}_{-3.5}$ &
      $11.8^{+7.4}_{-3.8}$ & $0.60^{+0.13}_{-0.23}$ & $0.21^{+0.03}_{-0.03}$ &
      $2.8^{+0.8}_{-0.4}$ & $0.9^{+0.4}_{-0.2}$ &
      $3.8^{+16.7}_{-3.3}\times10^{-11}$ & $1.3^{+0.2}_{-0.4}\times10^{-12}$
      \\[.8ex]

      \multicolumn{11}{l}{\textsc{tbnew(bbody+pcfabs(mekal)+pcfabs(mekal))}}
      \\[.4ex]
      & 1.00 & $5.1^{+12.3}_{-4.2}$ & $20.6^{+4.0}_{-4.0}$ &
      $~1.6^{+0.4}_{-0.4}$ & $0.99^{+0.01}_{-0.01}$ & $0.19^{+0.02}_{-0.03}$ &
      & & & ($2.7^{+0.4}_{-0.4}\times10^{-11}$)\tablefootmark{*}
      \\[0.8ex]

      & & & & $~6.1^{+4.1}_{-2.5}$ & $0.53^{+0.14}_{-0.21}$ & &
      $3.3^{+0.9}_{-0.6}$ & $1.2^{+0.6}_{-0.4}$ &
      $1.6^{+0.2}_{-0.2}\times10^{-11}$ & $7.6^{+0.4}_{-0.4}\times10^{-13}$
      \\[.4ex] \hline
    \end{tabular}
    \tablefoot{
      Fluxes have been determined via \textsc{cflux} within \textsc{xspec}.
      \tablefoottext{*}{The integrated flux of the low-temperature
        \textsc{mekal} component in this fit is highly sensitive even to small
        changes of temperature, abundance, and absorption, and should,
        therefore, be interpreted very carefully.}
    }
\end{table*}

%
%
\section{X-ray spectroscopy}
\label{sec:spectra}

  To derive the spectral parameters and the flux contribution of the system
  components and to gain information on the accretion geometry, we analyze the
  mean and the phase-resolved XMM-Newton spectra of QS~Tel. We fit the EPIC/pn
  and MOS2 spectra simultaneously in \textsc{xspec} v12.5
  \citep{arnaud:96,dorman:03}, combining black-body and \textsc{mekal} plasma
  emission components and twofold absorption terms:
  \textsc{tbnew}\footnote{Most recent version of \textsc{tbvarabs} in
    \textsc{xspec}. See
    \url{http://pulsar.sternwarte.uni-erlangen.de/wilms/research/tbabs/}.} to
  account for the galactic absorption,\linebreak using the cross-sections of
  \citet{verner:96a} and \citet{verner:96b} and the abundances of
  \citet{wilms:00}; and \textsc{pcfabs}, partially covering the plasma
  emission, to represent intrinsic photoelectric absorption. When determining
  \textsc{mekal} metal abundances, we refer to the results of
  \citet{asplund:09} from three-dimensional hydrodynamic modeling.

\subsection{The orbital mean spectrum}
\label{sec:meanspec}

  QS~Tel was in an intermediate high state of accretion during the XMM-Newton
  observation.  The spectral fits, thus, result in relatively low black-body
  and plasma temperatures, from which a rich metallic line spectrum
  arises. Our simplest fit consists of one black-body and one \textsc{mekal}
  component and the interstellar and intrinsic absorption terms. It implies a
  modest $\chi^2_\mathrm{red} = 1.22$ at $182$ degrees of freedom. The soft
  X-ray range ($0.1-0.4\,\mathrm{keV}$) is well described by the single black
  body at $kT_\mathrm{bb} = 21.1^{+3.5}_{-3.0}\,\mathrm{eV}$ and barely
  affected by interstellar absorption ($N_\mathrm{H, ISM} =
  3.3^{+5.9}_{-2.0}\times 10^{19}\,\mathrm{cm}^{-2}$). These results are in
  perfect agreement with the fit of \citet{rosen:96} to a ROSAT/PSPC spectrum,
  also taken during an intermediate high state, which yields $kT_\mathrm{bb} =
  20.5^{+1.5}_{-1.8}\,\mathrm{eV}$ and $N_\mathrm{H} = 3.3^{+1.0}_{-0.7}\times
  10^{19}\,\mathrm{cm}^{-2}$. The X-ray spectrum at energies above
  $0.4-0.5\,\mathrm{keV}$ is approximated by a \textsc{mekal} spectrum at a
  mean temperature of $kT_\textsc{mekal} = 2.2^{+0.4}_{-0.3}\,\mathrm{keV}$,
  comparable to the temperature of an only slightly absorbed bremsstrahlung
  component $kT_\mathrm{br} = 4.4\pm 0.9\,\mathrm{keV}$ that \citet{rosen:01}
  derive from an ASCA spectrum of QS~Tel. Additional plasma components
  decrease the $\chi^2_\mathrm{red}$ of our fits, indicating the need for
  multi-temperature models, but result in large error bars of the
  \textsc{mekal} temperatures and fluxes.

  We, therefore, test different temperature structures, using three
  alternative plasma models in \textsc{xspec}: \textsc{cemekl},
  \textsc{mkcflow}, and the accretion-flow model spectra that have been
  presented by \citet{traulsen:10} and are based on the calculations of
  \citet{fischer:01}. They fit the hard component slightly better than a
  single \textsc{mekal}. The effect is mainly visible from the Ca\,$4.6\,{\rm
    keV}$ and the Fe\,$6.7\,\mathrm{keV}$ lines. Anyhow, all the fits
  under-estimate the prominent calcium emission around
  $4.6\,\mathrm{keV}$. The element abundances resulting from the different
  fits deviate from each other by a factor of up to two, with a clear trend to
  subsolar values. The strong correlation between plasma temperature, mass
  accretion rate, and element abundance in the models cause the large
  systematic uncertainties when determining the chemical composition of the
  accretion stream. None of these approaches results in a significantly better
  $\chi^2_\mathrm{red}$ than two-\textsc{mekal} fits.

  \begin{figure}
    \centering
    \includegraphics[width=8.8cm]{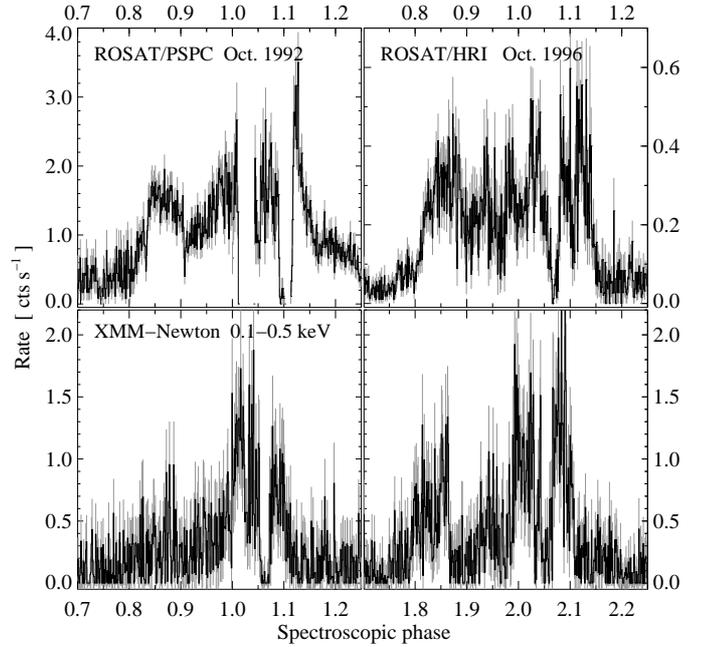}
    \caption{Soft bright-phase light curves of QS~Tel, binned into a time
      resolution of 10\,s. Upper panels: Archival ROSAT/PSPC and HRI
      data. Lower panels: EPIC/pn light curve in the 0.1$-$0.5\,keV band.}\vskip-5mm
    \label{fig:brightlc}%
  \end{figure}

\begin{figure}
    \centering
    \includegraphics[width=8.8cm]{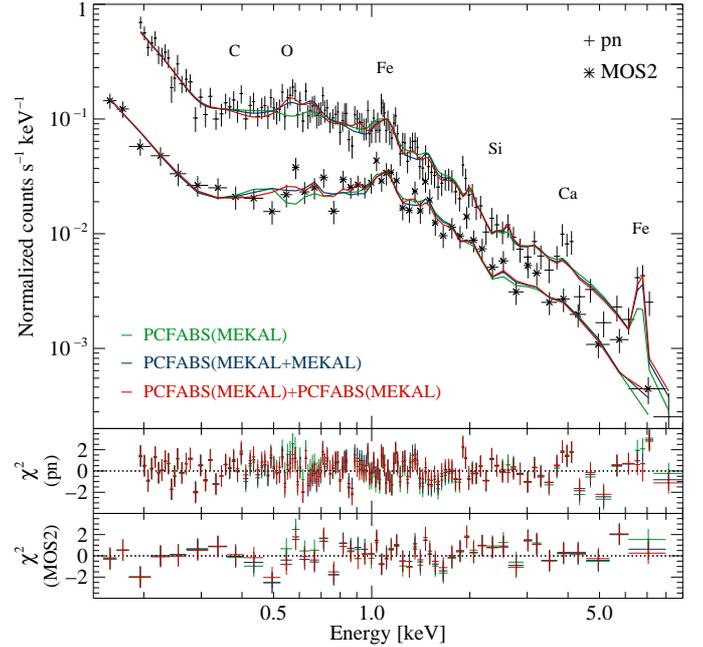}
    \caption{EPIC/pn and MOS2 spectra of QS~Tel and the \textsc{xspec} fits
      listed in Table~\ref{tab:xspecfits}. The models consist of a single
      black-body plus one and two partially absorbed \textsc{mekal}
      components, respectively. Spectral bins comprise a minimum of 20
      counts.}
    \label{spectrum}%
  \end{figure}

  The best fits are given by two multi-component models, each comprising two
  \textsc{mekal} components with independent temperatures and identical
  element abundances (see Table~\ref{tab:xspecfits} and Fig.~\ref{spectrum}).
  In the first model, we use one common absorption term for both the
  \textsc{mekal} components, assuming that they arise from the same emission
  region (\textsc{tbnew(bbody + pcfabs(mekal+mekal))}). It yields
  \textsc{mekal} temperatures of $kT_{\textsc{mekal},\rm low} =
  0.21^{+0.03}_{-0.03}\,{\rm keV}$ and $kT_{\textsc{mekal},\rm high} =
  2.8^{+0.8}_{-0.4}\,\mathrm{keV}$ at a mean element abundance of
  $0.9^{+0.4}_{-0.2}$ times the solar ($\chi^2_\mathrm{red}=1.06$ at 180
  d.\,o.\,f.).

  In the second model, we use an individual absorption term for each
  \textsc{mekal} component, assuming that they arise from different emission
  regions (\textsc{tbnew(bbody + pcfabs(mekal) + pcfabs(mekal))}). It infers
  the lowest $\chi^2_\mathrm{red}=1.00$ at 178 d.\,o.\,f. The low-temperature
  \textsc{mekal} component at $kT_{\textsc{mekal},\rm
    low}=0.19^{+0.02}_{-0.03}\,\mathrm{keV}$ is accompanied by intrinsic
  absorption of $N_\mathrm{H, intr}=1.6^{+3.7}_{-3.6}\times
  10^{22}\,\mathrm{cm}^{-2}$ with a high covering fraction close to one. The
  high-temperature component results in $kT_{\textsc{mekal},\rm
    high}=3.3^{+0.9}_{-0.6}\,\mathrm{keV}$ and $N_\mathrm{H,
    intr}=6.1^{+4.1}_{-2.5}\times 10^{22}\,\mathrm{cm}^{-2}$, which is
  partially covered by a factor of $0.5^{+0.1}_{-0.2}$. The absorption and
  temperature of the low-energy black-body component agree with the result of
  the \textsc{pcfabs(mekal+mekal)} fit within the confidence range.

\subsection{The X-ray bright and faint phases}
\label{sec:phase-res}

  The X-ray bright phases, defined by the on-off-behavior of the light curves
  in the softest energy range, cover the phase interval $\varphi =
  0.7-1.1$. We extract one mean bright- and one mean faint-phase spectrum for
  each EPIC instrument from the two orbital cycles covered and fit them with
  the one- and two-\textsc{mekal} spectra as described in
  Sect.~\ref{sec:meanspec}. Due to the smaller number of counts collected for
  the phase-resolved spectra, the uncertainties of the fit parameters are
  larger, in particular of the interstellar absorption term and the upper
  \textsc{mekal} temperature. Within these error bars, the resulting
  temperatures, abundances, and absorption terms can be considered as
  identical with the orbital means given in Table~\ref{tab:xspecfits} and,
  thus, as constant over the orbital cycle. In particular, we see no
  significant variation of the intrinsic absorption terms with the phase. The
  fluxes that we derive for the individual model components, though, are
  clearly phase-dependent and further discussed in Sect.~\ref{sec:excess}.

%
%
\section{Discussion}
\label{sec:discuss}

\subsection{Accretion geometry}
\label{sec:state}

  The most prominent features in the multi-wavelength light curves and the
  phase-resolved spectra can provide information on the accretion geometry of
  the system, useful for interpreting the spectral and flux components and for
  attributing them to the different emission regions. Former optical, UV, and
  soft X-ray observations have shown that QS~Tel changes between intervals of
  one-pole and two-pole accretion
  \citep{buckley:93,schwope:95,rosen:96,rosen:01}. \citet{schwope:95} identify
  the two accretion poles via two sets of cyclotron lines in the optical
  spectra. They suggest that one accretion pole with $B_1 \sim
  47\,\mathrm{MG}$, which is located near the connecting line to the
  secondary, mainly emits in the soft X-ray regime, and that a second
  accretion pole with $B_2 \sim 70-80\,\mathrm{MG}$ is optically bright. This
  also explains offsets in time between optically bright and EUV bright states
  as reported by \citet{rosen:01}. \citet{romero:04} confirm the existence of
  two accretion poles by high-state polarimetric observations. All the ROSAT
  soft X-ray \citep{schwope:95,rosen:96,rosen:01} and several EUV light curves
  \citep{buckley:93,rosen:01} are characterized by one pronounced maximum per
  orbital cycle. During various EUVE observations by
  \citet{rosen:96,rosen:01}, the light curves were double-peaked. These two
  maxima may indicate accretion onto the two, independently fed poles.

  Our soft X-ray light curves show the single maximum that can be explained by
  the orbital revolution of the main accretion region and are similar to the
  former ROSAT light curves. The detection of hard X-ray emission during the
  complete orbital cycle allows for two possible explanations of its nature:
  The main, soft X-ray bright accretion region is extended and undergoes a
  partial self-eclipse, or a significant fraction of the hard X-ray flux on
  the order of $50\,\%$ arises from the second, optically bright accretion
  region. We discuss these scenarios and potential photometric and
  spectroscopic signs of them in the following two Sections.

  \begin{figure*}
    \centering
    \includegraphics[width=18.4cm]{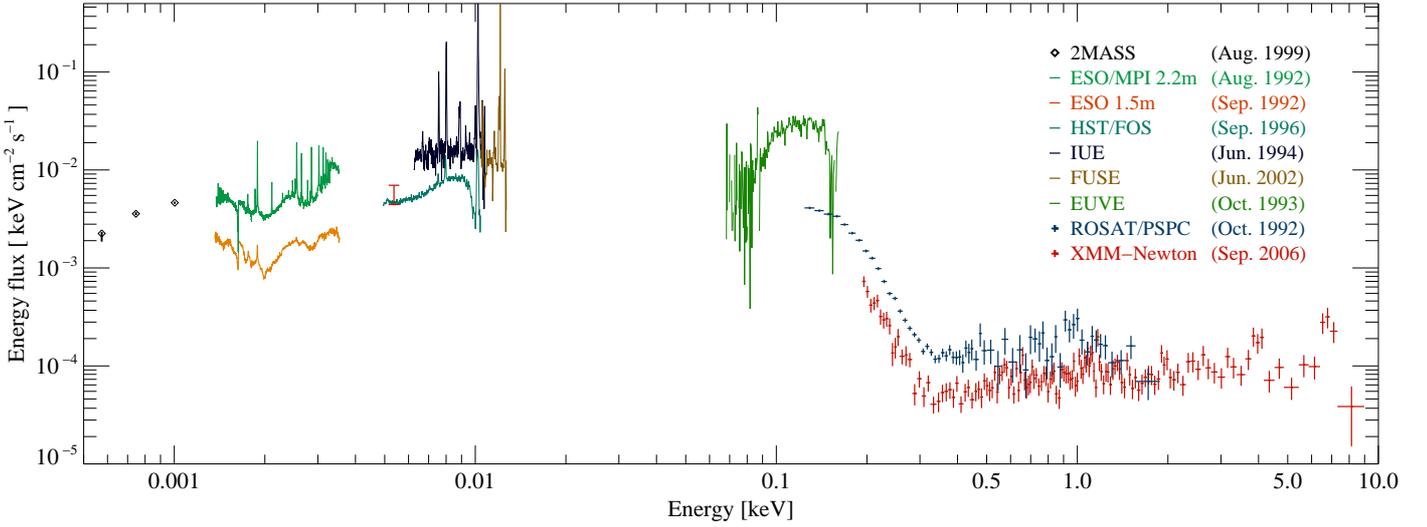}
    \caption{Spectral energy distribution of QS~Tel, including our XMM-Newton
      data and archival observations during different accretion states. The
      data point derived for the XMM-Newton Optical Monitor at 5.4\,eV
      (2310\,{\AA}) is shown as orbital minimum and maximum.}
    \label{fig:sed}%
  \end{figure*}

\subsubsection{A partially self-eclipsing main accretion region}
\label{sec:parteclipse}

  If the complete X-ray flux arises from the main pole, which almost faces the
  secondary, the faint-phase emission can be assigned to the uneclipsed outer
  parts of the accretion region. The accretion regions on the white-dwarf
  primaries in polars can show asymmetric and extended shapes, as investigated
  in detail for instance by \citet{wickramasinghe:91} and
  \citet{sirk:98}. Partial self-eclipses of such an extended accretion region
  have also been suggested by \citet{beuermann:87} for \object{EF~Eri} on the
  basis of Einstein light curves and by \citet{schwope:00} for \object{QQ~Vul}
  on the basis of polarimetric data. The soft X-ray emission of QS~Tel is
  reduced by a factor of about ten during the faint phase, possibly by the
  self-eclipse of the inner part of the extended accretion region, while the
  \textsc{mekal} flux diminishes by a factor between two and three. The
  different \textsc{mekal} components in the spectral fits are completely
  attributed to the main accretion column in this picture, and the temperature
  structure of this column deviates from the ones of typical multi-temperature
  models in \textsc{xspec} (see Sect.~\ref{sec:meanspec}). The similar
  \textsc{mekal} temperatures and partially covering absorption terms in the
  phase-resolved spectral\linebreak models, then, continuously reflect the
  main accretion column.

\subsubsection{Emission from the second accretion region}
\label{sec:sndregion}

  The (slightly variable) flux in the optical and ultraviolet wave bands can
  be attributed to the second, highly magnetic pole \citep{schwope:95}, which,
  thus, accreted at a low level at the epoch of our observations. If this
  accretion pole also shapes the faint-phase light curves and spectra, a
  cooler black-body-like component might arise from the region around the
  second accretion column, with its flux maximum possibly shifted to lower
  energies. In this picture, the two \textsc{mekal} components have their
  origin in the two accretion poles, and the different absorption terms in the
  \textsc{pcfabs(mekal) + pcfabs(mekal)} models imply that the plasma emission
  is absorbed to different degrees around the two accretion columns. The
  spectral parameters (black-body luminosities, normalizations of the
  \textsc{mkcflow} models) indicate a mass accretion rate of approximately
  $\dot{M}_\mathrm{faint} \sim 10^{-12}\,\mathrm{M_\odot}\mathrm{yr}^{-1}$ for
  the faint-phase and $\dot{M}_\mathrm{bright} \sim
  10^{-10}\,\mathrm{M_\odot}\mathrm{yr}^{-1}$ for the bright-phase fits. It is
  obvious that those models that rely on a typical temperature structure of a
  single accretion column cannot improve the fit quality compared to the
  two-\textsc{mekal} approaches, then (cf.\ Sect.~\ref{sec:meanspec}).
  Another two-pole accreting polar in which the second accretion region is
  probably seen during the X-ray faint phase is \object{BY~Cam}, one of the
  few asynchronous systems. \citet{ramsay:02} assign the flaring bright
  light-curve intervals of BY~Cam to a soft pole with 'blobby' accretion, and
  the low, hard faint-phase flux to the second pole, whose soft emission is
  shifted into the ultraviolet range. The authors state two alternative
  reasons for the lower temperatures in the accretion region, which would
  cause the wavelength shift: the region is less effectively heated due to a
  low accretion rate, or the accreted material is spread over a rather
  extended accretion area.

  For the interpretation of the spectral model components as emission of
  either one or two accretion pole(s), it may be instructive to evaluate the
  orbital modulation of their fluxes. Table~\ref{tab:fluxes} shows the
  phase-resolved bolometric fluxes that we derive from the two-\textsc{mekal}
  \textsc{xspec} models. For calculating the fluxes, temperatures and
  abundances are assumed to be constant over the binary cycle and set to their
  orbital means (Table~\ref{tab:xspecfits}). The flux of the low-temperature
  \textsc{mekal} component in the \textsc{pcfabs(mekal) + pcfabs(mekal)} fit
  stays essentially constant over the orbit, even though the large error bars
  have to be kept in mind. The high-temperature \textsc{mekal} flux changes
  clearly between bright and faint phase, but still to a lower degree than the
  black-body flux. This means: In case the two \textsc{mekal}s can be
  attributed to two accretion columns, then the black body and the hotter
  \textsc{mekal} have their origin in the primary, soft X-ray bright region,
  and the cooler \textsc{mekal} in the second, less active one. A signature of
  the second accretion pole in the XMM-Newton data, hence, is possible, but
  not definite.

\subsubsection{The soft X-ray light-curve dips}
\label{sec:dips}

  Two dips have been described in the light curves of QS~Tel at different
  wavelengths: A sharp dip recurs in the $\varphi = 0.0-0.1$ interval.  It has
  been found by \citet{rosen:96,rosen:01} and \citet{demartino:98} in X-ray,
  EUV, and UV data. A broader trough around $\varphi \sim 0.9$ appears
  occasionally in the X-ray light curves and varies in depth. It is well
  pronounced for example in the ROSAT/PSPC light curve profile of
  \citet{rosen:96}.

  Both dips can be identified in our X-ray light curves, which are similar to
  the ROSAT/PSPC and HRI light curves presented by
  \citet{rosen:96,rosen:01}. The broader trough around $\varphi \sim 0.9$
  appears both in the $0.1 - 0.5\,\mathrm{keV}$ and in the $0.5 -
  2.0\,\mathrm{keV}$ bands. Being variable and not strictly recurring, it
  could be an effect of random mass transfer variations, of structures in the
  accretion region, or an absorption feature, e.\,g. by an extended accretion
  column. \citet{rosen:96,rosen:01} report a similar pre-dip structure in a
  ROSAT/WFC light curve and consider absorption as the most plausible
  explanation.

  The narrow dip close to $\varphi \sim 0.07$ is mainly pronounced in the
  softest energy band and not seen in the optical light curves. Obviously a
  persisting feature, it also appears in the double-peaked EUVE light curves
  at similar orbital phase \citep{rosen:96,rosen:01}. Minor shifts of its
  phasing within the $\varphi \sim 0.0$ and 0.1 interval are found both in the
  ROSAT/HRI, PSPC, and WFC and in the EUVE light curves
  \citep[cf.][Fig.~\ref{fig:brightlc}]{rosen:96,rosen:01}. The dip is
  generally explained by the accretion stream eclipsing the X-ray bright
  accretion region. The short ingress and egress times and the hardness-ratio
  peak during the dip phase fit in this picture, although it is difficult to
  be reconciled with the interpretation of $\varphi = 0.0$ as inferior
  conjunction of the secondary (cf.\ Sect.~\ref{sec:photo}). Typically, the
  main accretion regions in polars are located at longitudes below $90\degr$
  \citep{cropper:88}, which places a potential stream dip in the $\varphi =
  0.75-1.0$ interval in this phase convention, less likely between $\varphi =
  0.05$ and $0.10$. A similar dip is seen in the EUVE light curves of
  \object{UZ~For} around orbital phase $0.9$ and could be due to stream
  absorption according to \citet{warren:95}. They point out that the influence
  of the mass accretion rate variations on the location of the threading
  region can explain why the phasing and the depth of the dips change between
  different orbital cycles. Similar flare-like activity characterizes our
  X-ray light curves (Sect.~\ref{sec:softlcs}) and former ROSAT light curves
  of QS~Tel \citep[stated e.\,g.\ by][]{rosen:96}, including the sharp
  dip. This, indeed, indicates $\dot{M}$ variations.

\begin{table}
    \caption{Unabsorbed bolometric fluxes of the phase-resolved
      two-\textsc{mekal} model components, using constant temperatures and
      abundances ($=$ orbital means) over the orbital cycle, and the
      corresponding black-body to \textsc{mekal} flux ratios.}
    \label{tab:fluxes} 
    \centering
    \begin{tabular}{*{2}{c@{\hskip2.9mm}}c@{~~}c@{~~}c}
      \hline\hline\\[-2.2ex]

      $F_\mathrm{bol}(\textsc{bbody})$ & $F_\mathrm{bol}(\textsc{mekal1})$ &
      $F_\mathrm{bol}(\textsc{mekal2})$ &
      $F_\mathrm{bb}/$ & $F_\mathrm{bb}/$ \\

      \multicolumn{3}{c}{[erg\,cm$^{-2}$\,s$^{-1}$]} & $F_\textsc{mekal}$ &
      $F_\textsc{mekal2}$ \\[0.3ex]

      \hline\\[-1.5ex]

      \multicolumn{5}{l}{\textsc{tbnew(bbody+pcfabs(mekal+mekal))}} \\[0.5ex]
      \multicolumn{5}{l}{Bright phase}\\[0.5ex]
      $1.2^{+1.6}_{-0.6}\times10^{-10}$ & $6.3^{+2.8}_{-2.5}\times10^{-13}$ &
      $1.8^{+0.4}_{-0.3}\times10^{-12}$ & $47^{+67}_{-23}$ & \\[1.2ex]

      \multicolumn{5}{l}{Faint phase}\\[0.5ex]
      $1.2^{+3.0}_{-0.9}\times10^{-11}$ & $2.1^{+1.3}_{-1.0}\times10^{-13}$ &
      $7.7^{+2.6}_{-1.8}\times10^{-13}$ & $12^{+31}_{-9}$ & \\[1.8ex]

      \multicolumn{5}{l}{\textsc{tbnew(bbody+pcfabs(mekal)+pcfabs(mekal))}}
      \\[0.5ex] \multicolumn{5}{l}{Bright phase}\\[0.5ex]
      $5.3^{+6.4}_{-3.0}\times10^{-11}$ & $1.7^{+5.4}_{-1.2}\times10^{-12}$ &
      $1.5^{+0.4}_{-0.3}\times10^{-12}$ & $17^{+20}_{-9}$ & $35^{+43}_{-20}$
      \\[1.2ex]

      \multicolumn{5}{l}{Faint phase}\\[0.5ex]
      $3.2^{+4.3}_{-1.2}\times10^{-12}$ & $1.6^{+0.6}_{-1.0}\times10^{-12}$ &
      $5.8^{+1.4}_{-1.0}\times10^{-13}$ & $1.5^{+2.0}_{-0.6}$ & $6^{+7}_{-2}$
      \\[0.6ex]

      \hline
    \end{tabular}
\end{table}

\subsection{The soft X-ray excess}
\label{sec:excess}

\subsubsection{Bolometric flux ratios}

  As the excess of soft over hard X-ray emission in polars is expected to
  increase with the magnetic field strength \citep{beuermann:94,ramsay:94},
  the high-field system QS~Tel can serve as a test case. Both the X-ray light
  curves and the spectra of QS~Tel indeed evince its very soft X-ray flux. The
  count rate decreases rapidly towards higher energies above
  $0.5\,\mathrm{keV}$. From our fits to the XMM-Newton spectra, we derive
  bolometric flux ratios of the black-body to the plasma components that are
  significantly larger than one. This finding is consistent with the flux
  ratios estimated by \citet{rosen:96} using EUVE and ROSAT
  data. \citet{ramsay:94} and \citet{ramsay:04}, on the other hand, determine
  luminosity ratios in the range between 1.2 and 10.6 from ROSAT spectra of
  QS~Tel, using a $30\,\mathrm{keV}$-bremsstrahlung model for the high-energy
  component.

  The actual values of the bolometric flux ratios during the X-ray bright
  phases depend on the spectral model and on the considered accretion
  scenario. Ratios of the total black-body to the total plasma flux correspond
  to the scenario of a partially eclipsed, extended main accretion region
  (Sect.~\ref{sec:parteclipse}). For the \textsc{pcfabs(mekal+mekal)} model,
  the integrated unabsorbed model fluxes, uncorrected for the viewing angle,
  give a ratio of $F_{\rm bb}/F_\textsc{mekal} = 47^{+67}_{-23}$ during the
  bright phases. The analogous approach for \textsc{pcfabs(mekal) +
    pcfabs(mekal)} results in $F_{\rm bb}/F_\textsc{mekal} = 17^{+20}_{-9}$,
  using the total flux of both \textsc{mekal} components. If the
  low-temperature \textsc{mekal} component is attributed to the second
  accretion region (scenario Sect.~\ref{sec:sndregion}), we have to compare
  the black-body and the high-temperature \textsc{mekal} component in order to
  derive the flux ratio of the primary accretion pole: $F_{\rm
    bb}/F_{\textsc{mekal},\mathrm{high}} = 35^{+43}_{-20}$. Thus, we can
  confirm a distinct soft X-ray excess of the primary accretion region in
  QS~Tel, in which the black-body component has its origin.

  Figure~\ref{fig:sed} compares the energy flux of our XMM-Newton and optical
  observations to archival multi-wavelength data. From our fits to the EPIC
  spectra, we derive a total integrated flux on the order of $F_\mathrm{bol}
  \sim 10^{-11}\,\mathrm{erg\,cm}^{-2}\mathrm{\,s}^{-1}$ and a bolometric
  model luminosity on the order of $L_\mathrm{bb} \sim 10^{31}
  (d/176\mathrm{pc})^2\,\mathrm{erg\,s}^{-1}$, using the distance determined
  by \citet{ak:07}. These results are two orders of magnitude lower than those
  determined for the EUVE high-state data by \citet{rosen:96}. The values that
  they derive from ROSAT spectra, also obtained during an intermediate high
  state of accretion, are similar to our XMM-Newton results. As a higher mass
  accretion rate is expected to affect the soft component mainly, our
  intermediate-state hardness and flux ratios can serve as lower limits for
  the high-state cases.

\subsubsection{The soft X-ray light curves}
\label{sec:softlcs}

  The hardness ratios, positive over large parts of the orbital cycle, reflect
  the little spectral contribution of events at energies above $0.5\,{\rm
    keV}$. A distinct soft X-ray excess is most probably connected to the
  `blobby` accretion scenario \citep{beuermann:04}, causing a flaring
  structure of the X-ray light curves. Despite the moderate count rate during
  the XMM-Newton pointing, individual flare events with a typical duration of
  $10-15\,\mathrm{s}$ appear in the soft X-ray light curves, mostly during the
  bright phases. These events can be assigned to separate filaments in the
  accretion stream permeating the white-dwarf atmosphere at a low rate. The
  patterns are comparable to the flares in the light curves of
  \object{V1309~Ori} \citep{schwarz:05}, which show -- at a considerably
  higher count rate -- clear evidence of single accretion blobs.

  The strong flickering appears to be present at all wavelengths from the
  optical down to soft X-rays
  \citep[e.\,g.][]{rosen:96,demartino:98,rosen:01}. \citet{warren:93} find
  flare-like events during a low state of accretion, which they explain as the
  potential impact of accretion blobs at a low mass transfer rate or as
  magnetic flares. This short-time variability provides further evidence of
  clumpy accretion events, as typical of soft X-ray dominated systems.

%
%
\section{Summary}

  At the epoch of our XMM-Newton observations, QS~Tel was in an intermediate
  high state of accretion, where most accretion activity was concentrated on
  the primary, soft X-ray bright pole. The soft X-ray light curves are
  characterized by a distinct bright phase, lasting about $40\,\%$ of the
  binary orbit. We identify the sharp dip in the soft X-ray light curve that
  has been seen in all previous ROSAT X-ray and various EUVE observations, and
  confirm that it is most likely due to an occultation of the soft X-ray
  emitting region by the accretion stream. The light curves in the hard X-ray,
  UV, and optical regime show little orbital modulation. The best fits to the
  X-ray spectra consist of one black-body and two absorbed plasma
  components. Whether these plasma components have their origin in different
  emission areas around the primary accretion pole or in independent accretion
  onto the two poles cannot be finally decided from our spectra. The
  modulation of the phase-resolved model fluxes indicates that the
  low-temperature plasma component might represent the second, optically
  bright accretion pole. The bolometric flux ratios of the soft (black-body)
  and the hard (\textsc{mekal}) components are strongly model dependent and in
  the range of 15 to about 120 during the bright phases. Together with the
  high short-term variability of the soft X-ray light curves, this finding
  establishes QS~Tel as a distinctly soft X-ray dominated, 'blobby' accreting
  system.

%
%
\begin{acknowledgements}
  This research has been granted by DLR under project numbers 50\,OR\,0501 and
  50\,OR\,0807. We have obtained the ROSAT/PSPC and HRI data from the High
  Energy Astrophysics Science Archive Research Center (HEASARC), provided by
  NASA's Goddard Space Flight Center, and the EUVE, FUSE, IUE, and HST spectra
  from the Multimission Archive at the Space Telescope Science Institute
  (MAST).

\end{acknowledgements}

%
%
\bibliographystyle{aa}
\bibliography{16352}

\end{document}